\documentclass{elsart5p}

\usepackage{graphicx}

\usepackage{amssymb}

\begin{document}

\begin{frontmatter}



\title{Weak quasistatic magnetism in the frustrated Kondo lattice Pr$_2$Ir$_2$O$_7$}



\author[UCR]{D.~E. MacLaughlin\corauthref{cor}}
\corauth[cor]{tel. +1-951-827-5344,
Fax: +1-951-827-4529,
email: macl@physics.ucr.edu}
\author[ISSP]{Y. Ohta}
\author[ISSP]{Y. Machida}
\author[ISSP]{S. Nakatsuji}
\author[McMaster]{G.~M. Luke}
\author[KU]{K. Ishida}
\author[RHH]{R.~H. Heffner}
\author[UCR]{Lei Shu\thanksref{UCSD}}
\author[CSULA]{O. O. Bernal}


\address[UCR]{Department of Physics and Astronomy, University of California, Riverside, California 92521, U.S.A.}
\address[ISSP]{Institute for Solid State Physics, University of Tokyo, Kashiwa 277-8581, Japan}
\address[McMaster]{Department of Physics and Astronomy, McMaster University, Hamilton, Ontario, Canada L8S 4M1}
\address[KU]{Department of Physics, Graduate School of Science, Kyoto University, Kyoto 606-8502, Japan}
\address[RHH]{Los Alamos National Laboratory, Los Alamos, New Mexico 87545, U.S.A.}
\address[CSULA]{Department of Physics and Astronomy, California State University, Los Angeles, California 90032, U.S.A.}
\thanks[UCSD]{Present address: Department of Physics and Institute for Pure and Applied Physical Sciences, University of California, San Diego, La Jolla, California 92093, U.S.A.}


\begin{abstract}
Muon spin relaxation experiments have been performed in 
the pyrochlore iridate Pr$_2$Ir$_2$O$_7$ for temperatures in the range 0.025--250~K\@. Kubo-Toyabe relaxation functions are observed up to $\gtrsim 200$~K, indicating static magnetism over this temperature range. The $T \rightarrow 0$ static muon spin relaxation rate~$\Delta(0) \approx 8~\mu\mathrm{s}^{-1}$ implies a weak quasistatic moment (${\sim}0.1\mu_B$)\@. The temperature dependence of $\Delta$ is highly non-mean-field-like, decreasing smoothly by orders of magnitude but remaining nonzero for $T < T_f$. The data rule out ordering of the full Pr$^{3+}$ CEF ground-state moment (3.0~$\mu_B$) down to 0.025~K\@. The weak static magnetism is most likely due to hyperfine-enhanced $^{141}$Pr nuclear magnetism. The dynamic relaxation rate~$\lambda$ increases markedly below $\sim$20~K, probably due to slowing down of spin fluctuations in the spin-liquid state. At low temperatures $\lambda$ is strong and temperature-independent, indicative of a high density of low-lying spin excitations as is common in frustrated antiferromagnets. 
\end{abstract}


\end{frontmatter}

\section{Introduction}
\label{sec:introduction}
A wide variety of conducting and magnetic behavior, including metal-insulator transitions and strongly frustrated magnetism, is found in the $R_2$Ir$_2$O$_7$ ($R = 4f$ ion) series of pyrochlore-structure iridates~\cite{YaMa01}. Pr$_2$Ir$_2$O$_7$ is a metallic compound in which inelastic neutron scattering experiments~\cite{MNTT05} indicate a well-isolated Pr$^{3+}$ $\Gamma_3$ magnetic doublet CEF ground state. Transport properties show behavior similar to the Kondo effect, suggesting partial screening of the moments, even below the RKKY interaction temperature~$T_{\mathrm{RKKY}} \approx 20$~K~\cite{NMMT06}; geometrical frustration was suggested as a mechanism for suppression of magnetic order. The magnetic susceptibility exhibits an unusual $-\ln T$ temperature dependence over more than a decade of temperature below $\sim$2~K, and below $\sim$0.12~K a history-dependent magnetization gives evidence for partial spin freezing.

This paper reports results of preliminary muon spin relaxation ($\mu$SR) experiments in polycrystalline samples of Pr$_2$Ir$_2$O$_7$ for temperatures in the range 0.025--250~K\@. Kubo-Toyabe (K-T) muon spin relaxation functions~\cite{HUIN79} are observed in low applied fields, indicating weak but well-resolved quasistatic magnetism up to relatively high temperatures $\sim 150$~K\@. The rapid component of the K-T relaxation reflects a distribution of quasistatic components~$\langle \mathbf{B}_{\rm loc}\rangle$ of the local fields~$\mathbf{B}_{\rm loc}$ at muon sites. Its relaxation rate $\sim 8~\mu\mbox{s}^{-1}$ at low temperatures indicates a small quasistatic moment, of the order of $0.1\mu_B$\@. The slower dynamic relaxation is due to thermal fluctuations of $\mathbf{B}_{\rm loc}$. Strong but smooth variations of the quasistatic and dynamic relaxation rates with temperature were observed at all temperatures, with no evidence for a phase transition below 150~K but with large deviations from simple mean-field-like behavior. The data are consistent with quasistatic hyperfine-enhanced $^{141}$Pr nuclear magnetism~\cite{Blea73,AbBl83}, which raises the question of the origin of the nonmagnetic Pr$^{3+}$ ground state required for this phenomenon.

The $\mu$SR experiments were carried out in a $^3$He-$^4$He dilution refrigerator and a $^4$He gas-flow cryostat at the M15 and M20B beam lines, respectively, at TRIUMF, Vancouver, Canada. Two powder samples of Pr$_2$Ir$_2$O$_7$, prepared as described previously~\cite{MNTT05} and labeled PI8E and PI40I, were studied over the overlapping temperature ranges~0.025--20~K and 2.1--250~K, respectively. Longitudinal fields of a few millitesla were applied to prevent muon precession in small stray fields. 

Data from both samples are well fit by a damped static Gaussian K-T function~\cite{HUIN79}
\begin{equation} P(t) = \exp(-\lambda t)\,G_L(H_L,\Delta t) \,, \label{eq:dampedstatic} \end{equation}
where $G_L(H_L,\Delta t)$ is the static Gaussian K-T function in applied longitudinal field~$H_L$~\cite{HUIN79} and is characterized by a relaxation rate~$\Delta$. This form is appropriate to the case of concentrated local moments with random orientations, the dipolar fields of which to a good approximation produce Gaussian distributions of the components of $\langle \mathbf{B}_{\rm loc}\rangle$, with zero mean and rms width~$\Delta/\gamma_\mu$, where $\gamma_\mu$ is the muon gyromagnetic ratio. The damping factor~$\exp(-\lambda t)$ models the situation where $\mathbf{B}_{\rm loc}(t)$ is the superposition of a quasistatic component~$\langle \mathbf{B}_{\rm loc}\rangle$ and a thermally fluctuating component~$\delta\mathbf{B}_{\rm loc}(t)$. Fits obtained using the dynamic K-T function~\cite{HUIN79}, which describes fluctuations of the entire local field~$\mathbf{B}_{\rm loc}(t)$, were not as good ($\chi^2 \approx 1.3$) as the damped Gaussian fits ($\chi^2 \approx 1.05$). Fits using damped static or dynamic Lorentzian K-T functions, which describe relaxation by dilute spins~\cite{UYHS85}, were even worse than the dynamic K-T fits. 

\begin{figure}[ht]
\centering
\includegraphics[width=\columnwidth,clip]{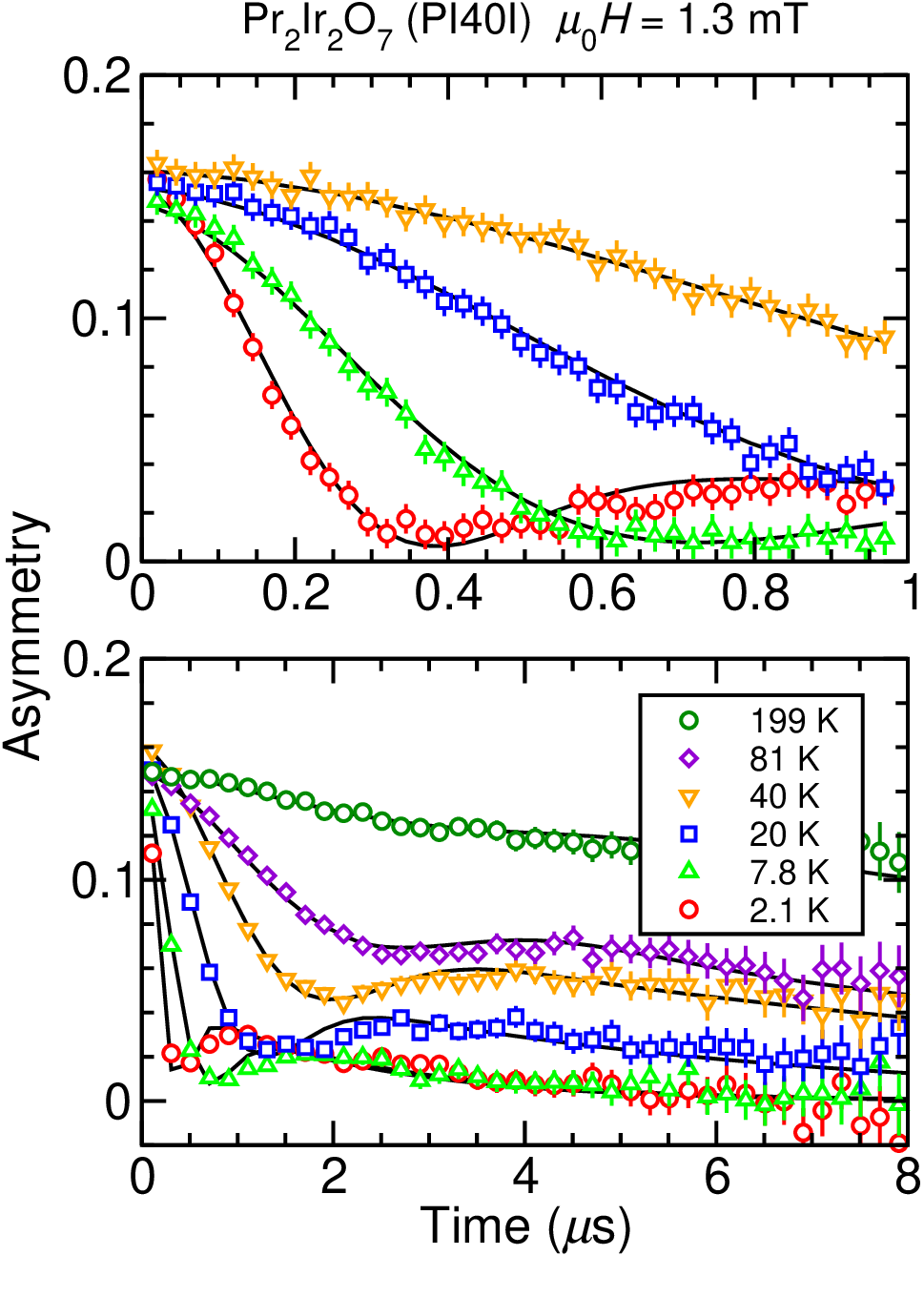}
\caption{Representative early- and late-time muon spin relaxation functions in Pr$_2$Ir$_2$O$_7$, sample PI40I, $2.1\mbox{ K} \le T \le 199\mbox{ K}$.}
\label{fig:PI40I-asy}
\end{figure}

Figure~\ref{fig:PI40I-asy} gives representative muon spin relaxation data and fits for the sample PI40I\@. The two-component K-T form of the 
relaxation functions is evident. The base-temperature value~$\Delta(0.025\,\mathrm{K}) = 7.6(3)~\mu\mbox{s}^{-1}$ corresponds to a rms quasistatic local field $\langle \mathbf{B}_{\rm loc}\rangle_{\mathrm{rms}} = \Delta/\gamma_\mu = 8.9(4)$~mT\@. The relation between this field and the value of the moments that produce it depends on the muon stopping site or sites, which are not known. The moment value is of the order of $0.1~\mu_B$, assuming concentrated moments and dipolar interactions. The Pr$^{3+}$ CEF ground-state moment is ${\sim}3.0~\mu_B$~\cite{MNTT05}, so freezing of the entire Pr spin system is ruled out. The observed dynamic relaxation rate~$\lambda \sim 10^6~\mu\mathrm{s}^{-1}$ gives an upper bound on the fluctuation rate of the quasistatic moments.

\begin{figure}[ht]
\centering
\includegraphics[width=\columnwidth,clip]{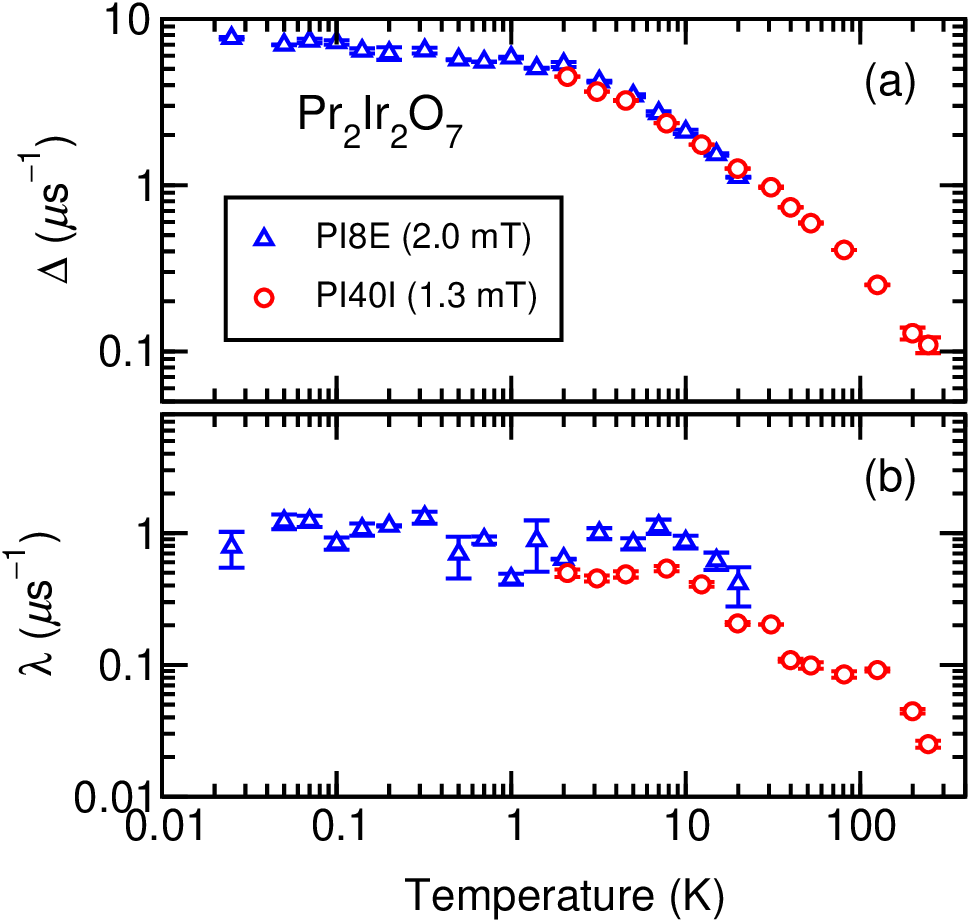}
\caption{Temperature dependencies of (a)~static relaxation rate~$\Delta$ and (b)~dynamic relaxation rate~$\lambda$ in Pr$_2$Ir$_2$O$_7$, samples PI8E (triangles) and PI40I (circles).}
\label{fig:Deltalambda}
\end{figure}

The temperature dependencies of the relaxation rates~$\Delta$ and $\lambda$ are given in Figs.~\ref{fig:Deltalambda}(a) and \ref{fig:Deltalambda}(b), respectively. The values of $\Delta$ for the two samples agree extremely well in
the overlap region, indicating that the static magnetism is intrinsic. The difference of a factor of $\sim$2 in $\lambda$ between the samples is discussed below. Both $\Delta$ and $\lambda$ exhibit smooth temperature dependencies. The temperature dependence of $\Delta$ is highly non-mean-field-like: $\Delta(T)$ is nonzero but very small compared to $\Delta(0) = \Delta(T{\rightarrow}0)$ over most of the temperature range below $\sim$100~K\@. At ${\sim}$4~K $\Delta(T)$ has already decreased to $0.5\Delta(0)$, and at ${\sim}$80~K to $0.05\Delta(0)$. There is evidence for more abrupt decreases in both $\Delta$ and $\lambda$ above 200~K [Fig.~\ref{fig:Deltalambda}(a)] that could be due to a phase transition, but more likely arise from the onset of thermally-activated muon diffusion.

We argue that the weak magnetism is most likely due to hyperfine-enhanced~\cite{Blea73,AbBl83} $^{141}$Pr nuclear moments. Hyperfine-enhanced nuclear magnetism is a Van Vleck-like response of a non-Kramers nonmagnetic 4$f$ CEF ground state to the nuclear moment due to the atomic hyperfine coupling. The resulting admixture of 4$f$ and nuclear wave functions enhances the effective nuclear moment by an enhancement factor~$K = a_{4f}\chi_{\mathrm{mol}}$, where $\chi_{\mathrm{mol}}$ is the molar 4$f$ susceptibility and $a_{4f}$ is the atomic hyperfine coupling constant. Thus $K$ is the chemical shift, which can be large in 4$f$ ions in metals due to the strong hyperfine interaction and large Van Vleck susceptibility. Values of $K \gtrsim 20$ have been observed in Pr-based systems with nonmagnetic ground states~\cite{MHNC00,SMAT07}. 

\begin{figure}[ht]
\centering
\includegraphics[width=\columnwidth,clip]{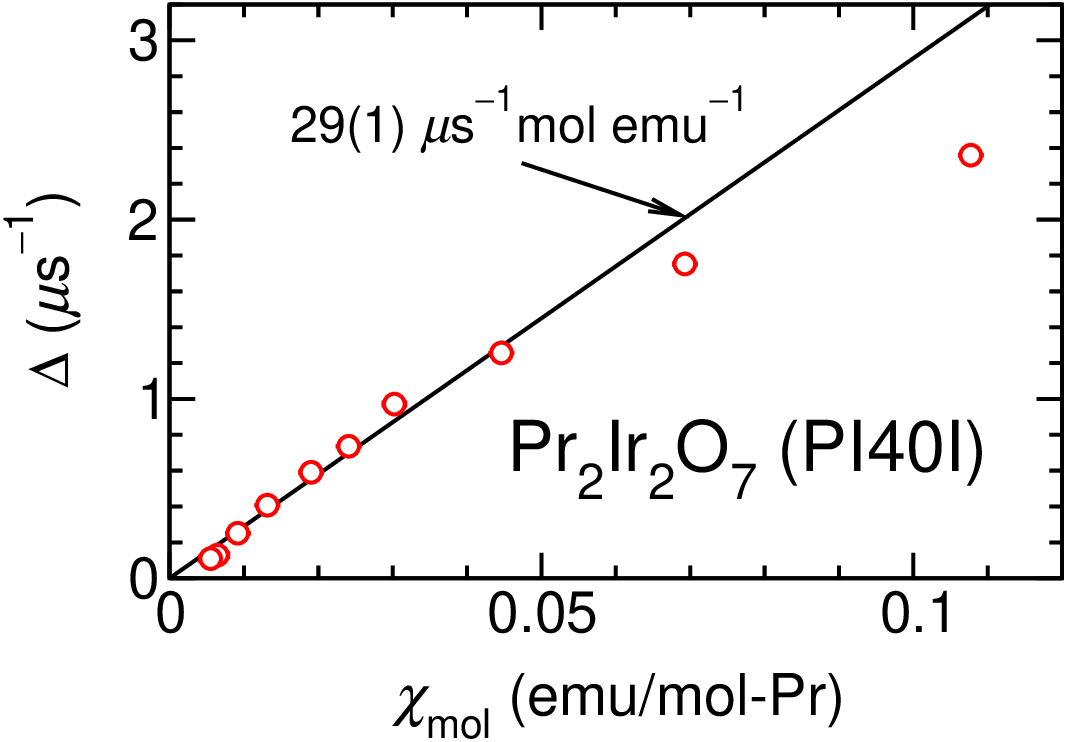}
\caption{Dependence of static relaxation rate~$\Delta$ on bulk molar susceptibility~$\chi_{\mathrm{mol}}$ in Pr$_2$Ir$_2$O$_7$, sample PI40I.}
\label{fig:Deltavschi}
\end{figure}

Hyperfine-enhanced $^{141}$Pr nuclear moments couple to the muon spin via their correspondingly enhanced dipolar fields. Their contribution to the static muon relaxation rate is, therefore~\cite{Blea73},
\begin{equation} \Delta = (1 + K) \Delta_0 \,, \quad \mbox{i.e.,} \quad d\Delta/d\chi_{\mathrm{mol}} = a_{4f}\Delta_0 \,, 
\label{eq:hfenhancedDelta}
\end{equation}
where $\Delta_0$ is the unenhanced value of the static $^{141}$Pr contribution. Thus a plot of $\Delta(T)$ vs $\chi_{\mathrm{mol}}(T)$, with temperature an implicit parameter, should be a straight line with slope~$a_{4f}\Delta_0$. In Fig.~\ref{fig:Deltavschi} it can be seen that $\Delta(\chi_{\mathrm{mol}})$ is 
roughly linear at high temperatures (small $\chi_{\mathrm{mol}}$), with a slope $\approx 29~\mu\mathrm{s}^{-1}~\mathrm{mol~emu}^{-1}$, which from Eq.~(\ref{eq:hfenhancedDelta}) and $a_{4f} = 187.7~\mathrm{mol~emu}^{-1}$ gives $\Delta_0 \approx 0.16\ \mu\mathrm{s}^{-1}$. If the muon stopping site or sites is known $\Delta_0$ can be calculated, but unfortunately this not the case in Pr$_2$Ir$_2$O$_7$. Nevertheless $\Delta_0/\gamma_\mu \approx 180~\mu$T is a typical value for abundant nuclei with nuclear gyromagnetic ratios~$\gamma_{\mathrm{nuc}}/2\pi \sim 10$~MHz/T, as for $^{141}$Pr.

This result is compelling evidence for the hyperfine-enhancement scenario, but a number of questions remain to be answered. The Pr$^{3+}$ CEF ground state is a non-Kramers magnetic doublet. In a nearly free ion the $^{141}$Pr moment would be rapidly relaxed and unable to yield the quasistatic muon local field evidenced by the data of Fig.~\ref{fig:PI40I-asy}~\cite{HUIN79}. As noted above, the $\mu$SR results put an upper bound on this fluctuation rate of ${\sim}10^6~\mu\mathrm{s}^{-1}$. 

The deviation of $\Delta(\chi_{\mathrm{mol}})$ from a straight line at low temperatures (large $\chi_{\mathrm{mol}}$, Fig.~\ref{fig:Deltavschi}) might, however, signal modification of near-neighbor Pr$^{3+}$ CEF ground states by the positive muon electric charge, as observed in PrNi$_5$~\cite{FAGG95b} and PrIn$_3$~\cite{TAGG97}. If strong enough, this modification could produce a nonmagnetic singlet Pr$^{3+}$ ground state, thereby quenching the $^{141}$Pr nuclear relaxation and establishing the conditions necessary for hyperfine nuclear enhancement. 

It is hard to estimate the size of muon-induced crystal-field effects. CEF changes of the order of 10~K were found in other systems~\cite{FAGG95b,TAGG97}, and Van Duijn et al.~\cite{vDKHA05} reported evidence from specific heat and neutron scattering experiments for disorder-induced splittings of this order in the related pyrochlores~Pr$_{2-x}$Bi$_x$Ru$_2$O$_7$. Our observation of quasistatic hyperfine-enhanced $^{141}$Pr nuclear magnetism in Pr$_{2}$Ir$_2$O$_7$ up to $\sim$200~K would be surprising if the muon-induced splitting were as small as this. An intriguing alternative is that in Pr$_{2}$Ir$_2$O$_7$ the nonmagnetic ground state required for hyperfine nuclear enhancement is a many-body effect, i.e., due to robust singlet spin-liquid formation rather than CEF splitting. 

For completeness we consider an electronic origin for the weak magnetism. The data are consistent with a frozen-moment state without long-range magnetic order (consistent with nuclear spins well above any nuclear ordering temperature). The temperature dependence of $\Delta$ is very smooth and shows no anomalies below $\sim$150~K; in particular, the absence of an increase in $\Delta$ below $T_{\mathrm{RKKY}}$ is microscopic evidence that the Pr$^{3+}$ moments do not freeze at this temperature. There is no signature in either $\Delta(T)$ or $\lambda(T)$ of freezing for $T \lesssim 200$~K, and no signature whatsoever in the bulk susceptibility.

Freezing of the full CEF ground-state moment of the entire Pr$^{3+}$ spin system is ruled out as noted above; furthermore, not many Pr-based materials order magnetically at temperatures $\sim$100~K\@. A frozen Pr-spin concentration of a few percent with the full moment would yield the correct magnitude of $\Delta(0)$ but should also give rise to Lorentzian K-T relaxation, whereas the observed relaxation has the Gaussian K-T form~\cite{UYHS85} appropriate to concentrated moments. Weak-moment freezing of Ir$^{4+}$ spins ($S = 1/2$) has been observed at $T_f \approx 150$--170~K in the insulating pyrochlore iridate~Y$_2$Ir$_2$O$_7$~\cite{YaMa01,TWH01,FuMa02,ASKS03} with, however, an abrupt onset that is not observed in Pr$_2$Ir$_2$O$_7$. Furthermore, to our knowledge Ir$^{4+}$ spin freezing has never been reported in a metallic iridate.
  
The dynamic relaxation rate~$\lambda$ increases by an order of magnitude with decreasing temperature below $\sim$20~K for both samples. This temperature is close to $T_{\mathrm{RKKY}}$, and is also the region where transport and thermodynamic properties indicate the onset of a low-temperature spin-liquid state~\cite{NMMT06}. The increase of $\lambda$ is unlikely to be due to Pr$^{3+}$ (or Ir$^{4+}$) spin freezing, since there is no corresponding increase of $\Delta$. It suggests either slowing down of Pr$^{3+}$ spin fluctuations during a broad crossover into the spin-liquid regime, or a related change in hyperfine-enhanced $^{141}$Pr nuclear spin dynamics. 

As noted above, the temperature dependence of $\lambda$ exhibits a difference between the low- and high-temperature measurements. In a frustrated magnet dependence of properties on small sample differences is not unexpected. Alternatively, a systematic error may arise in the data taken below 2~K because of the significant fraction of muons stopped in the cold finger of the dilution refrigerator. This fraction must be determined from the fits, and is strongly correlated with other fit parameters such as the rate. In spite of this uncertainty it is clear that $\lambda(T)$ remains roughly constant down to 0.025~K, with a value in the range~0.5--$1~\mu\mbox{s}^{-1}$. 

Persistent dynamic muon spin relaxation down to the lowest temperatures of measurement is a common feature of frustrated antiferromagnets~\cite{Koji01,MOBB07}, and has been taken as evidence for a large and possibly singular density of low-energy spin excitations. Temperature-independent relaxation would not be expected from $^{141}$Pr nuclear spin fluctuations due to enhanced $^{141}$Pr-$^{141}$Pr spin-spin interactions, since these would reflect the temperature dependence of $\chi_{\mathrm{mol}}$~\cite{Blea73,SMAT07}. They would also result in dynamic K-T relaxation rather than the observed damped static form [Eq.~(\ref{eq:dampedstatic})].
 
To summarize, $\mu$SR experiments reveal weak but well-resolved quasistatic magnetism in Pr$_2$Ir$_2$O$_7$ over a wide temperature range. For a number of reasons it is doubtful that this magnetism is electronic in nature. Hyperfine-enhanced nuclear magnetism seems most likely, because it has the correct semiquantitative relation with the magnetic susceptibility, but the origin of the required nonmagnetic ground state is unclear; muon-induced Pr$^{3+}$ CEF splitting and many-body Pr$^{3+}$ singlet formation are candidate mechanisms. Future work is clearly necessary to elucidate this situation.

We are grateful to the \mbox{TRIUMF} Centre for Molecular and Materials Science, where these experiments were carried out, for technical assistance, and to R.~F. Kiefl for useful discussions. This work was supported by the U.S. NSF, Grants~0422674 (Riverside) and 0604015 (Los Angeles), by Grants-in-Aid for Scientific Research from JSPS, and by a Grant-in-Aid for Scientific Research on Priority Areas (19052003) (Tokyo).




\end{document}